\documentclass[%
 reprint,
 amsmath,amssymb,
 aps,showpacs,prl
]{revtex4-1}

\usepackage{graphicx}
\usepackage{dcolumn}
\usepackage{bm}
\newcommand{\bb}{\bibitem}

\begin{document}

\preprint{APS/123-QED}

\title{Masslesslike minimal subtraction for massive scalar field theory}
\author{M. M. Leite}%
 \email{marcelo.mleite@ufpe.br}
\affiliation{Laborat\'orio de F\'\i sica Te\'orica e Computacional, Departamento de F\'\i sica,\\ Universidade Federal de Pernambuco,\\
50670-901, Recife, PE, Brazil}


\begin{abstract}
 {\it We introduce the simplest minimal subtraction method for massive $\lambda \phi^{4}$ field theory with $O(N)$ internal symmetry, which resembles the same method applied to massless fields by using two steps. First, the utilization of the partial-$p$ operation in every diagram of the two-point vertex part in order to separate it into a sum of squared mass and external momentum, respectively, with different coefficients. Then, the loop integral which is the coefficient of the quadratic mass  can be solved entirely in terms of the mass, no longer depending upon the external momentum, using the {\it parametric dissociation transform}. It consists in the choice of a certain set of fixed values of Feynman parameters replaced inside the remaining loop integral after solving the internal subdiagrams. We check the results in the diagrammatic computation of critical exponents at least up to two-loop order using a flat metric with Euclidean signature.}   
\end{abstract}
\pacs{11.10.-z; 03.70.+k; 64.60.F-}

\maketitle
\section{1. Introduction}
\par With the invention  of the renormalization group \cite{W,WK}, scalar field theories have been explored in a wide range of physical situations. Some instances include elementary particle physics, unveiling important properties of the standard model \cite{H} as well as semianalytical multi-loop calculations with massive particles in the context of deep inelastic scattering \cite{RR}. There are also examples in gravitation and cosmology. (It was realized long ago that classical gravitation can be formulated in terms of scalar fields \cite{BD}). Quantum gravity effects can be obtained at least at one-loop order through the linearization of the metric tensor in the Einstein-Hilbert action interacting with a free massive scalar field, resulting  in a vertex with two scalars and the graviton. Resummation techniques in the infrared region can be utilized in order to improve the ultraviolet regime of the quantum theory \cite{Ward}. Furthermore, scalar, spinors and vector massless fields described in a cosmological background with a dilaton scalar field playing the role of the cosmological constant for especial values of its vacuum expectation value and coupling with the quantum fields  is an example of fine tuning in cosmology which generates the masses of those fields\cite{Tamarit}.
\par One of the most important applications of renormalization group ideas is perhaps the perturbative computations of critical properties of many body critical systems undergoing phase transitions \cite{BLZ}. Minimal subtraction schemes \cite{tHV,LA} are particularly simple in dealing with massless fields, but become somewhat involved in the treatment of massive fields. Is it possible to enunciate a minimal subtraction technique for massive fields which captures the same essential  pattern of the simplest version of its massless counterpart?
\par In this letter we commence to shed light on this issue by devising such a method for massive scalar fields in a $\lambda \phi^{4}$ theory.  It is appropriate to call it "masslesslike" massive minimal subtraction since it resembles that for massless fields.  It requires a minimal number of diagrams,  precluding diagrams which include tadpole insertions (unlike in the BPHZ method \cite{BPHZ}). With this shortcut, we show that all primitively divergent vertex parts (see Refs.\cite{BLZ,Amit}) that can be renormalized multiplicatively are rendered finite utilizing our method. Since the perturbative formulation of spinor and vector fields  ultimately reduce to the computation of Feynman integrals of scalar fields \cite{S},  this perfected scheme of minimal subtraction has the potential of application in several instances of renormalized perturbative computations of quantum massive fields in the high energy (ultraviolet or simply UV) regime. 
\par We begin with the bare Lagrangian density in $d$-dimensional flat space (Euclidean or Minkowski spacetime with index $\nu$) for scalar fields written as
\begin{equation}\label{1}
\mathcal{L} = \frac{1}{2} \partial_{\nu} \phi \partial^{\nu} \phi + \frac{1}{2} \mu_{0}^{2} \phi^{2} 
+ \frac{\lambda}{4!} (\phi^{2})^{2},
\end{equation} 
where $\mu_{0}$ and $\lambda$ are the bare mass and coupling constant, respectively. We omitted the index corresponding to the $O(N)$ internal symmetry since they appear only as $N$-dependent coefficients in the diagrammatic expansion. The one-particle irreducible ($1PI$) primitively divergent vertex parts are $\Gamma^{(2)}(p, \mu_{0}, \lambda, \Lambda)$, 
$\Gamma^{(4)}(p_{i}, \mu_{0}, \lambda, \Lambda)$ and the composite one $\Gamma^{(2,1)}(p_{1}, p_{2}; Q,\mu_{0}, \lambda, \Lambda)$ . The parameter $\Lambda$ is the cutoff which characterizes the bare theory under consideration \cite{IIM}.  The $p_{i}$ in the argument of the several vertex parts stand for external momenta, whereas $Q$  is the momentum of the inserted composite operator.  Note that a vertex part with an arbitrary number $N$ of external legs and of composite operators $L$,  is represented by $\Gamma^{(N,L)}(p_{1},,... p_{N}; Q_{1},..., Q_{L},\mu_{0}, \lambda, \Lambda)$.
\par Let us summarize some basic facts. The starting point is to define a three-loop bare mass $\mu^{2}$= $\Gamma^{(2)} (p=0, \mu_{0}, \lambda, \Lambda) = \mu_{0}^{2} + O(\lambda)$. Next we write $\mu_{0}^{2}= \mu^{2} - O(\lambda)$ and replace it  in all the vertex parts. Consequently, the vertex parts now depend only upon $\mu$ in their arguments. This procedure has the virtue of eliminating {\it all} diagrams including tadpole insertions inside {\it all} vertex parts at arbitrary loop order.  For the aim we have in mind, we expand $\Gamma^{(2)}$ up to three-loop order, whereas 
$\Gamma^{(4)}$ and $\Gamma^{(2,1)}$ are expanded up to two-loop level. The diagrams that are left in $\Gamma^{(2)}$ have the peculiarity that they must be subtracted from their value at $p=0$ order by order at the loop expansion (see the conventions in \cite{CL}). Without loss of generality, we shall consider a Euclidean metric that will be useful for our purposes in what follows. The results, however, are valid (with minor modifications) to include Minkowski spacetime. Henceforth  we shall drop the cutoff from the arguments of all vertex parts: dimensional regularization of the divergent integrals will be expressed as poles in $\epsilon=4-d$ throughout.       
\par The aforementioned bare primitively divergent vertex parts have the following perturbative expansions:
\begin{subequations}\label{2}
\begin{eqnarray}
&& \Gamma^{(2)} (p,\mu,\lambda) =  p^{2} + \mu^{2} - \frac{\lambda^{2}}{6} \frac{(N+2)}{3} 
[I_{3}(p,\mu) - I_{3}(0,\mu)] \nonumber \\ 
&& + \frac{\lambda^{3}}{4} \frac{(N+2)(N+8)}{27} [I_{5}(p,\mu)  -  I_{5}(0,\mu)], \label{2a}\\
&& \Gamma^{(4)} (p_{i},\mu,\lambda) = \lambda - \frac{\lambda^{2}}{2} \frac{(N+8)}{9} 
[I_{2}(p_{1} + p_{2}, \mu) \nonumber\\ 
&& + 2perms. ] + \frac{\lambda^{3}}{4} \frac{(N^2 + 6N + 20)}{27} [ I_{2}^{2} (p_{1} + p_{2},\mu)\nonumber\\
&&  + 2 perms. ] + \frac{\lambda^{3}}{2} \frac{(5N+22)}{27} [ I_{4} (p_{i}, \mu) + 5 perms. ],\label{2b}\\
&&  \Gamma^{(2,1)} (p_{1}, p_{2}; Q, \mu,\lambda) = 1 - \lambda \frac{(N+2)}{18} [I_{2}(p_{1} + p_{2}, \mu) \nonumber\\
&& + 2perms.] + \lambda^{2} \frac{(N+2)^{2}}{108}  [ I_{2}^{2} (p_{1}+p_{2}, \mu) \nonumber\\
&& + 2 perms. ] + \lambda^{2} \frac{(N+2)}{36} [ I_{4}(p_{1}, p_{2}, Q, \mu) + 5 perms. ]\label{2c}
\end{eqnarray}.
\end{subequations}
\par The integrals $I_{2}$ and $I_{4}$ are not particularly important to the manipulations we are going to make. They are given by
\begin{eqnarray}
&I_{2}(P, \mu) = \int \frac{d^{d} q}{(q^{2} + \mu^{2})[(q+P)^{2} + \mu^{2}]},\nonumber\\
&I_{4} (p_{i}, \mu) = \int \frac{d^{d} q_{1} d^{d} q_{2}}{(q_{1}^{2} + \mu^{2})(q_{2}^{2} + \mu^{2})} \nonumber\\
& \; \times  \frac{1}{[(P-q_{1})^{2} + \mu^{2}][(q_{1} - q_{2} + p_{3})^{2} + \mu^{2}]}. \nonumber
\end{eqnarray}
\par We simply give their expressions in terms of $\epsilon$ as $I_{2} (P,\mu)=\frac{\mu^{- \epsilon}}{\epsilon} \Bigl[1 -\frac{\epsilon}{2} (1 + L(P,\mu))\Bigr]$ and $I_{4} =\frac{\mu^{- 2 \epsilon}}{2 \epsilon^{2}} \Bigl[ 1 - \epsilon (L(P,\mu) + \frac{1}{2}) \Bigr]$, where $P= (p_{1} + p_{2}, p_{1} + p_{3}, p_{2} + p_{3})$ and $L(P,\mu) = \int_{0}^{1} dx ln\Bigl[\frac{P^{2}}{\mu^{2}} x(1-x) + 1 \Bigr]$. 
\par The key ingredient is the manipulation of the two- and three-loop Feynman integrals $I_{3}$ and $I_{5}$ of the two-point vertex part. They correspond to the following expressions:
\begin{subequations}\label{3}
\begin{eqnarray}
&I_{3}(p,\mu) = \int \frac{d^{d} q_{1} d^{d} q_{2}}{(q_{1}^{2} + \mu^{2})(q_{2}^{2} + \mu^{2})[(q_{1} + q_{2} + p)^{2} + \mu^{2}]},\label{3a}\\
&I_{5}(p,\mu) = \int \frac{d^{d} q_{1} d^{d} q_{2} d^ {d} q_{3}}{(q_{1}^{2} + \mu^{2})(q_{2}^{2} + \mu^{2})(q_{3}^2 + \mu^ {2})[(q_{1} + q_{2} + p)^{2} + \mu^{2}]}\nonumber\\
& \times \; \frac{1}{[(q_{1} + q_{3} + p)^{2} + \mu^{2}]},\label{3b}
\end{eqnarray}
\end{subequations}
\par  We now apply  the "partial-$p$" operation in the two-point vertex part $\Gamma^{(2)}$ in the two and three-loop diagrams. Our paradigmatic example to be discussed here is the two-loop diagrams which we are left with. We apply it in the form $\frac{1}{2d} [\frac{\partial q_{1}^{\mu}}{\partial q_{1}^{\mu}} 
+ \frac{\partial q_{2}^{\mu}}{\partial q_{2}^{\mu}}]$, where $q_{i}$ are the loop momentum. For the three-loop graphs we have to use the partial-$p$ inside the integrand in a different form, involving all loop momenta (with $3d$ in the denominator of the operation).    
\par Consider $I_{3}$. First apply the "partial-$p$" operation in the form given above. A preliminary result is 
\begin{subequations}\label{4}
\begin{eqnarray}
&I_{3}(p,\mu) = -\frac{1}{(d-3)} \Bigl[3 \mu^ {2} A(p,\mu)+ B(p,\mu) \Bigr], \label{4a}\\
&A(p,\mu) = \int  \frac{d^{d} q_{1} d^{d} q_{2}}{(q_{1}^{2} + \mu^{2})^{2} (q_{2}^{2} + \mu^{2})[(q_{1} + q_{2} + p)^{2} + \mu^{2}]}, \label{4b}\\
&B(p,\mu) = \int  \frac{d^{d} q_{1} d^{d} q_{2}  p. (q_{1} +q_{2} +p)}{(q_{1}^{2} + \mu^{2}) 
(q_{2}^{2} + \mu^{2})[(q_{1} + q_{2} + p)^{2} + \mu^{2}]^{2}}.\label{4c}
\end{eqnarray}
\end{subequations}
\par The issue is how to get rid of the $p$-dependence of the integral $A(p,\mu)$, since its dependence is there to stay. This can be done  using the "parametric dissociation transform" 
$(PDT)$ to be described now. Observe that 
$I_{2}(q_{1} +p, \mu)$ appears as a subdiagram of $I_{3}(p, \mu)$. Solving for this internal bubble (using the conventions from ref. \cite{Amit}), utilizing another Feynman parameter and making the continuation $\epsilon = 4 - d$,  we find
\begin{eqnarray}\label{5}
& A(p,\mu) = \frac{1}{2} \Gamma(2-\frac{\epsilon}{2}) \Gamma(2+\frac{\epsilon}{2}) 
\int_{0}^{1} dx [x(1-x)]^{-\frac{\epsilon}{2}}  \int_{0}^{1} dy \nonumber\\ 
& \int \frac{y^{\frac{\epsilon}{2} -1} (1-y) d^{d} q_{1}}{\Bigl[q_{1}^{2} + 2q_{1}.py + p^{2}y + \mu^{2}[1-y+ \frac{y}{x(1-x)}]\Bigr]^{2 +\frac{\epsilon}{2}}}.
\end{eqnarray}
\par We start with $PDT$ by setting $y=0$ inside the momentum loop integral, since this has the virtue of eliminating all dependence on the external momentum $p$ and what is left is a function of $\mu$ only. Note that this value of the parameter corresponds to the endpoint singularity of the $y$ integration, which will maximize the loop momentum contribution as far as the $y$-dependence is concerned keeping, therefore, the correct pole structure of the integral $A(p,\mu$). We then find
\begin{equation}\label{6}
A(p,\mu)= \frac{\mu^{-2 \epsilon}}{2 \epsilon^{2}} \Bigl[ 1 - \frac{\epsilon}{2} + \epsilon^{2} \Bigl(\frac{\pi^ {2}}{12} + \frac{1}{2} \Bigr) \Bigr]
\end{equation}
\par If we employ the same $PDT$ in $I_{3}(0,\mu)$, then we consistently obtain $(A(p,\mu)-A(0,\mu))_{PDT} =0$. This asset is now available and make the connection with the simplest version of the minimal subtraction scheme for massless fields \cite{Amit}. Without loss of generality, we shall omit $\mu$ from the arguments of the several diagrams, since it is obvious from the beginning that this parameter is there. Now the remaining integral can be shown to be given by  
$B(p, \mu)= \frac{\mu^{-2 \epsilon} p^{2}}{8 \epsilon} \Bigl[1-\frac{3 \epsilon}{4} - 2 \epsilon L_{3}(p,\mu) \Bigr]$, which implies, using the parametric dissociation transform, in the expression 
\begin{subequations}\label{7}
\begin{eqnarray} 
&& I_{3} (p) - I_{3}(0) = -\frac{\mu^{-2 \epsilon} p^{2}}{8 \epsilon} 
\Bigl[1+\frac{\epsilon}{4} - 2 \epsilon L_{3}(p,\mu)\Bigr], \label{7a}\\
&& L_{3}(p,\mu) = \int_{0}^{1} dx dy (1-y)ln \Bigl[\frac{p^{2}}{\mu^{2}} y(1-y) \nonumber\\
&& + 1 - y +\frac{y}{x(1-x)}\Bigr]. \label{7b}
\end{eqnarray}
\end{subequations} 
\par The same procedure can be applied to $I_{5}$. Using the partial-$p$ operation, 
the integral $I_{5}(p)$ now reads
\begin{subequations}\label{8}
\begin{eqnarray}
&I_{5}(p) = -\frac{2}{(3d-10)} \Bigl[\mu^{2} (C_{1}(p,\mu) + 4 C_{2}(p,\mu)) \nonumber\\
& \qquad + 2 D(p,\mu)\Bigr],\label{8a}\\
&C_{1}(p,\mu) = \int \frac{d^{d} q_{1} d^{d} q_{2} d^ {d} q_{3}}{(q_{1}^{2} + \mu^{2})^{2}(q_{2}^{2} + \mu^{2})(q_{3}^2 + \mu^ {2})[(q_{1} + q_{2} + p)^{2} + \mu^{2}]}\nonumber\\
& \times \; \frac{1}{[(q_{1} + q_{3} + p)^{2} + \mu^{2}]},\label{8b}\\
&C_{2}(p,\mu) = \int \frac{d^{d} q_{1} d^{d} q_{2} d^ {d} q_{3}}{(q_{1}^{2} + \mu^{2})(q_{2}^{2} + \mu^{2})^{2} (q_{3}^2 + \mu^ {2})[(q_{1} + q_{2} + p)^{2} + \mu^{2}]}\nonumber\\
& \times \; \frac{1}{[(q_{1} + q_{3} + p)^{2} + \mu^{2}]},\label{8c}\\
&D(p,\mu) =  \int  \frac{d^{d} q_{1} d^{d} q_{2} d^{d} q_{3} p.(q_{1} + q_{2} + p)}{(q_{1}^{2} + \mu^{2}) (q_{2}^{2} + \mu^{2})(q_{3}^{2} + \mu^{2})} \nonumber\\
&\qquad \frac{1}{[(q_{1} + q_{2} + p)^{2} + \mu^{2}]^{2}[(q_{1} + q_{3} + p)^{2} + \mu^{2}]}.\label{8d}
\end{eqnarray}
\end{subequations}
Let us analyze $C_{1}(p,\mu)$ ($C_{2}(p,\mu)$ can be studied analogously). Integrating independently the two internal bubbles (each one turns out to be $I_{2}(q_{1}+p,\mu)$) and employing extra Feynman parameters, we are left with
\begin{eqnarray}\label{9}
&C_{1}(p,\mu)= \frac{\Gamma(2+\epsilon)}{\Gamma^{2}(\frac{\epsilon}{2})} \int_{0}^{1} [x(1-x)]^{-\frac{\epsilon}{2}} dx \int_{0}^{1} [y(1-y)]^{-\frac{\epsilon}{2}} dy\nonumber\\
&\int_{0}^{1} z^{\frac{\epsilon}{2}+1} (1-z)^{\frac{\epsilon}{2} -1} dz \int_{0}^{1} w 
(1-w)^{\frac{\epsilon}{2} -1} dw \int d^{d}q_{1}\nonumber\\
&\frac{1}{\Bigl[q_{1}^{2} + 2p. q_{1}(1-z w) + p^{2}(1-z w) + \mu^{2}\Bigl[z w +\frac{z(1-w)}{x(1-x)} + \frac{1-z}{y(1-y)} \Bigr] \Bigr]^{2+\epsilon}}
\end{eqnarray}
\par The $PDT$ implementation in the loop integral follows the same principle: replace into the momentum integral the values $z=w=1$ (endpoint singularities of their parametric integrals). The principle is the same for arbitrary loops. We then find 
$C_{1}(p,\mu)=\frac{\mu^{-3\epsilon}}{3 \epsilon^{3}} \Bigl[1 -\frac{\epsilon}{2}  + \frac{\epsilon^{2}}{4}(3 + 5 \psi^{'}(1)) \Bigr]$ where $\psi^{'}(z)=\frac{d^{2} ln \Gamma(z)}{dz^{2}}$.  The same set of fixed values can be employed in the computation of $C_{2}(p,\mu)$. Indeed, at each loop order, the same set of fixed parameters can be used to solve as many integrals of the $C_{i}(p,\mu)$ type as there are homotopically 
different sets of diagrams after the partial-$p$ operation is applied, resulting in momentum independent results as shown above.  After performing the integral $C_{2}(p,\mu)$ we find $(C_{1}(p,\mu) + 4C_{2}(p,\mu)) = -\frac{\mu^{-3 \epsilon}}{3 \epsilon^{3}} \Bigl[1 - \frac{5 \epsilon}{2} + \frac{\epsilon^{2}}{4}(11 + \frac{\pi^{2}}{6}) \Bigr]$. When subtracted from the value at $p=0$, this combination does not contribute. The integral $D(p,\mu)$ is easy to compute.
By keeping in mind these set of steps we get to the expression:
$[I_{5} (p) - I_{5} (0)] = -\frac{p^{2} \mu^{-3 \epsilon}}{6 \epsilon^{2}} \Bigl[ 1+ \frac{\epsilon}{2} -3 \epsilon L_{3}(p, \mu) \Bigr].$
\par  A comment is in order.  The method of ref.  \cite{CL} requires an extra subtraction of the standard minimal subtraction procedure,  specifically at the two point vertex part.  After the extra subtraction,  it turns out that it no longer satisfies the nonperturbative Callan-Symanzik ($CS$) equation.  The same vertex part produces composite operators which are not identical to the standard ones {\it because of the extra subtraction}.  Consequently,  the scaling limit of the $CS$ equation in the ultraviolet regime for the massive theory is never attained in the context of that work.  It is opportune to point out that,  as it is going to be shown below,  the standard argument of minimal subtraction apllies in a straightforward manner to the present method. This is in stark contrast with the problems plaguing the method aforementioned.

\section{3. Application}
\par Let us express the bare and renormalized coupling contants in terms of dimensionless ones through $\lambda=u_{0} \mu^{\epsilon}$ and $g=u \mu^{\epsilon}$, respectively. In the four-point vertex part, whose first term is put in evidence, the remaining loop terms only depend on 
$\mu$ through the logarithmic integrations.  In the other vertex parts,  
those definitions suppress the overall dependence on 
$\mu$ in every diagram. This dependence only occurs through the parametric logarithmic integrals as well,  resulting in a bare perturbative expansions in terms of the bare dimensionless coupling constant $u_{0}$. The renormalized vertex parts built out of the primitively divergent can be written as 
$\Gamma_{R}^{(2)}(p,m,u)=Z_{\phi} \Gamma^{(2)}(p,\mu,u_{0})$, 
$\Gamma_{R}^{(4)}(p_{i},m,u)= Z_{\phi}^{2} \Gamma^{(4)}(p_{i},\mu,u_{0})$ and 
$\Gamma_{R}^{(2,1)}(p_{1},p_{2}; Q,m,u)= Z_{\phi} Z_{\phi^{2}} \Gamma^{(2,1)}(p_{1},p_{2};Q,\mu,u_{0})$, respectively 
($\bar{Z}_{\phi^{2}} \equiv Z_{\phi} Z_{\phi^{2}})$. Therefore, any vertex part which is multiplicatively renormalizable satisfies the equation 
$\Gamma_{R}^{(N,L)}(p_{i};Q_{j},m,u) = Z_{\phi}^{\frac{N}{2}} Z_{\phi^{2}}^{L} \Gamma^{(N,L)}(p_{i};Q_{j},\mu,u_{0})$ ($i=1,...,N$; $j=1,...,L$), where the normalization functions $Z_{\phi}$ and $Z_{\phi^{2}}$ are determined entirely from the finiteness of the renormalized vertex parts obtained from the primitively divergent vertex parts.
The expansions
\begin{subequations}\label{10}
\begin{eqnarray} 
&& u_{0} = u(1+a_{1} u +a_{2} u^{2}), \label{10a} \\
&& Z_{\phi} = 1+b_{2} u^{2} +b_{3} u^{3}, \label{10b} \\
&& \bar{Z}_{\phi^{2}} = 1 + c_{1} u +c_{2} u^{3}, \label{10c} 
\end{eqnarray}
\end{subequations}
in terms of the renormalized coupling constant will suffice for our program. 
\par Let us state the renormalization by starting with the two-point function up to three-loops, but only the  computation of $b_{2}$ will be made explicit.  The bare vertex function can be written as as  $\Gamma^{(2)} = p^{2} + \mu^{2} -B_{2} u_{0}^{2} + B_{3} u_{0}^{3}$.   Note that $B_{2} = \mu^{2\epsilon} [I_{3}(p) -  I_{3}(0)]$ and $B_{3} = \mu^{3\epsilon} [I_{5}(p) -  I_{5}(0)] $.  Forget for the time being the last term which will be important in the computation of $b_{3}$. Multiplicative renormalizability implies that 
$\Gamma_{R}^{(2)}(p,m,u) = Z_{\phi} \Gamma^{(2)}(p, \mu, u_{0}) = p^{2} + Z_{\phi} \mu^{2} + (b_{2}p^{2} -B_{2}) u^{2}$ is finite (at this order $u_{0}^{2}=u^{2}$).  Since we analyze the two-point vertex part at three-loop order, we {\it define the renormalized mass at third order in perturbation theory} as $m^{2} = Z_{\phi} \mu^{2}$.  Had we worked at $l$-loop order, we would have defined the renormalized mass at $l$-loop order in the same manner, with $Z_{\phi}$ computed at $l$-loop order. (Without loss of generality we restrict ourselves only up to three-loop level.)  This is similar to what happens in the massless theory: there are no tadpoles if we impose that the renormalized mass is zero to all orders in perturbation theory, which follows from the same equation at arbitrary loop order when setting $\mu=0$. By demanding that the renormalized two-point vertex part to be finite at two-loop level, we find $b_{2}= -\frac{(N+2)}{144 \epsilon}$. 
\par Now, $\Gamma_{R}^{(4)}(p_{i},m,u) = Z_{\phi}^{2} \Gamma^{(4)}(p_{i}, \mu, u_{0})$ and the logarithmic integrals whose coefficients contain poles in $\epsilon$ cancel out in the perturbative expansion after we expand the dimensionless bare coupling constant in terms of the dimensionless renormalized one (in the computation of $a_{2}$). This yields $a_{1} = \frac{(N+8)}{6 \epsilon}, a_{2}= \frac{(N+8)^{2}}{36 \epsilon^{2}} -\frac{(3N+14)}{24 \epsilon}$. Using a similar reasoning for the composite vertex, namely that the explicit cancellation of the logarithmic integrals takes place in a similar manner as occurred with the four-point function just discussed,  and including the computation of three-loop contribution belonging to the two-point function, we can write the normalization functions in the form
\begin{subequations}\label{11}
\begin{eqnarray}
&&\bar{Z}_{\phi^{2}}= 1+ \frac{(N+2)}{6 \epsilon} u + \Bigl[\frac{(N+2)(N+5)}{36 \epsilon^{2}} \nonumber\\
&& \quad \quad - \frac{(N+2)}{24 \epsilon}\Bigr] u^{2},\label{11a} \\ 
&& Z_{\phi} = 1 -\frac{(N+2)}{144 \epsilon} u^{2} - \Bigl[\frac{(N+2)(N+8)}{1296 \epsilon^{2}} \nonumber \\
&& \quad \quad + \frac{N+2)(N+8)}{5184 \epsilon}\Bigr] u^{3}. \label{11b} 
\end{eqnarray}
\end{subequations}
\par  The Euclidean metric chosen can be utilized to check universality in critical phenomena through the  diagrammatic calculation of critical exponents using the  $CS$ framework \cite{CS,C,V}, since in the Lagrangian (1), the mass is proportional to $|T-T_{C}|$. Therefore, the temperature of the system is away from the critical temperature $T_{C}$ characterizing the phase transition in the present setting. 
\par By considering the vertex parts in terms of the dimensionless coupling constants and applying the operator $m \frac{\partial}{\partial m}$ on $\Gamma_{R}^{(N,L)}$, we find $\Bigl[m \frac{\partial}{\partial m} + \beta(u) \frac{\partial}{\partial u} -\frac{N}{2} \gamma_{\phi}(u) + L \gamma_{\phi^{2}}\Bigr] \Gamma_{R}^{(N,L)}(p_{i};Q_{j},m,u)= \frac{(2-\gamma_{\phi}(u))m^{2}}{\bar{Z}_{\phi^{2}}}  \Gamma_{R}^{(N,L+1)}(p_{i};Q_{j},0,m,u)$. In the original argument, the right-hand side ($rhs$) of this equation was obtained using normalization conditions for the renormalized mass by setting $N=2,L=0$ \cite{Amit}. Since the renormalized mass in our method is not obtained from a fixed value of external momenta as in normalization conditions, it results in the appearance of the term $(\bar{Z}_{\phi^{2}})^{-1}$ in the $CS$ equation. Then, by demanding independence of the renormalization scheme, we simply have to set the tree-level value  $\bar{Z}_{\phi^{2}}=1$ within the context of our method and the $CS$ equation turns out to be the same either using normalization conditions or the present minimal subtraction method, namely
\begin{eqnarray}
&\Bigl[m \frac{\partial}{\partial m} + \beta(u) \frac{\partial}{\partial u} -\frac{N}{2} \gamma_{\phi}(u) + L \gamma_{\phi^{2}}\Bigr] \Gamma_{R}^{(N,L)}(p_{i};Q_{j},m,u)= \nonumber\\
&(2-\gamma_{\phi}(u))m^{2}  \Gamma_{R}^{(N,L+1)}(p_{i};Q_{j},0,m,u).
\end{eqnarray} 
After that, in the UV regime, the right-hand side ($rhs$) of this equation is neglected in comparison with the ($lhs$) \cite{Wein}, and scaling is valid \cite{Amit}.  
\par The Wilson function $\beta(u) = m \Big(\frac{\partial u}{\partial m})_{\lambda} 
= -m \frac{(\frac{\partial \lambda}{\partial m})_{u}}{(\frac{\partial \lambda}{\partial u})_{m}}$  that can be rewritten as
\begin{eqnarray}
&& \beta(u)= -\epsilon \left(\frac{\partial lnu_{0}}{\partial u}\right)^{-1} = u \Bigl[-\epsilon + \frac{(N+8)}{6} u \nonumber\\
&& \qquad \qquad - \frac{(3N+14)}{12} u^{2}\Bigr]
\end{eqnarray}
 has a nontrivial (repulsive) UV fixed point 
($\beta(u_{\infty})=0$), namely $u_{\infty} = \frac{6 \epsilon}{(N+8)}\Bigl[1 + \frac{3(3N+14)}{(N+8)^{2}} \epsilon \Bigr]$. The function $\gamma_{\phi}(u)= \beta(u) \frac{\partial ln Z_{\phi}}{\partial u}= u\Bigl[\frac{(N+2)}{72} u - \frac{(N+2)(N+8)}{1728} u^{2}\Bigr]$ when computed at the fixed point, yields the (anomalous dimension of the field) exponent 
$\eta(\epsilon)$ up to three-loop order, namely
\begin{equation}
\eta= \frac{(N+2)\epsilon^{2}}{2(N+8)^{2}}\Bigl[1+\Bigl(\frac{6(3N+14)}{(N+8)^{2}} -\frac{1}{4}\Bigr)\epsilon \Bigr].
\end{equation}
\par Moreover, the function $\bar{\gamma}_{\phi^{2}} (u)= - \beta(u) \frac{\partial ln \bar{Z}_{\phi^{2}}}{\partial u}= \frac{(N+2)}{6}u\Bigl[1-\frac{u}{2} \Bigr]$ when evaluated at the fixed point along with the identity $\nu^{-1}= 2 - \bar{\gamma}_{\phi^{2}} (u_{\infty}) - \eta$ produces the correct correlation length exponent $\nu(\epsilon)$ as 
\begin{equation}
\nu= \frac{1}{2} + \frac{(N+2)}{4(N+8)} \epsilon + \frac{(N+2)(N^{2} + 23N + 60)}{8(N+8)^{3}} \epsilon^{2}. 
\end{equation}
The details will be reported elsewhere.
\section{4. Discussion and Conclusions}
\par It turns out  that the present method  has several advantages over all previously massive renormalization schemes. First, in comparison with normalization conditions \cite{BLZ2} it is much simpler. Second, when compared with the $BPHZ$ minimal subtraction method \cite{BP,Hepp,Z}, a minimal number of diagrams is required. Third, in the case studied here, the renormalized mass receives no "radiative corrections" but is defined by the product of the three-loop bare mass, which is an arbitrary parameter, multiplied by the normalization function $Z_{\phi}$. 
\par Since the renormalized mass is not obtained from normalization conditions but defined as discussed above, the appearance of $\bar{Z}_{\phi^{2}}$ in the $CS$ equation is a residual effect of this definition and can be neutralized by setting its tree-level value $\bar{Z}_{\phi^{2}}=1$. This results in the "covariance" of the $CS$ equation by using either normalization conditions or the present minimal subtraction scheme. Since the $CS$ equation is a nonperturbative tool, valid order by order in perturbation theory, this general feature will be maintained in higher-loop orders. 
\par The nontrivial determination of the renormalization functions is almost the same as in massless theories. Moreover, the method keeps the pole structure of the coefficients of the squared bare mass in two-point vertex parts, but they do not contribute due to the especial properties of the perturbative expansion chosen herein.
\par The present framework might be able to address the renormalization of perturbative expansions of quantum massive scalar fields in particle physics.  For instance, in the scalar setor of the Higgs Doublet Model \cite{Lee}. A recent study of minimal subtraction renormalization \cite{BOR}  beyond one-loop level would be feasible within the context of our method.  Moreover, it might offer a simple alternative to tackling massive scalar field renormalization in an external potential \cite{FSM}.
\par In critical phenomena, systems confined in a parallel plate geometry represented by massive fields can now be treated within this minimal subtraction generalizing the treatment in the massless scheme for periodic and antiperiodic boundary conditions \cite{BL} for the field. They remain to be investigated in the massive theory and with more general boundary conditions \cite{SBL}. 
\par Curiously, the cancellation of tadpoles in the massive formulation of Ref. \cite{SBL} with a more complicated internal tensor structure is analogous to the perturbative expansion of the vertex parts discussed in the present proposal. Indeed, the finite-size ($FS$) effect is implemented as an internal symmetry as $(O(N) \times (FS))$ and works in the same manner as presented here, but using normalization conditions. It will be interesting to apply the present formalism in the same problem and see whether it needs any adaptation. Its application in the renormalization of generic Lifshitz competing systems \cite{L} is left for future research.
\par We would like to acknowledge partial support from CAPES (Brazilian agency) through the PROEX Program 534/2018 grant number 23038.003382/2018-39.

\end{document}